# Real-time full bandwidth measurement of spectral noise in supercontinuum generation

B. Wetzel [1], A. Stefani [1], L. Larger [1], P. A. Lacourt [1], J. M. Merolla [1], T. Sylvestre [1], A. Kudlinski [2], A. Mussot [2], G. Genty [3], F. Dias [4], J. M. Dudley [1*]

1. Institut FEMTO-ST, UMR 6174 CNRS-Université de Franche-Comté, Besançon, France
2. PhLAM/IRCICA  CNRS-Université Lille 1, USR 3380/UMR 8523, F-59655 Villeneuve d'Ascq, France
3. Department of Physics, Tampere University of Technology, Tampere, Finland
4. School of Mathematical Sciences, University College Dublin, Belfield, Dublin 4, Ireland

\* Corresponding Author

J. M. Dudley

john.dudley@univ-fcomte.fr

Tel: +33381666494

Fax: +33381666423

**ABSTRACT**

The ability to measure real-time fluctuations of ultrashort pulses propagating in optical fiber has provided significant insights into fundamental dynamical effects such as modulation instability and the formation of frequency-shifting rogue wave solitons. We report here a detailed study of real-time fluctuations across the full bandwidth of a fiber supercontinuum which directly reveals the significant variation in measured noise statistics across the spectrum, and which allows us to study correlations between widely separated spectral components. For two different propagation distances corresponding to the onset phase of spectral broadening and the fully-developed supercontinuum, we measure real time noise across the supercontinuum bandwidth, and we quantify the supercontinuum noise using statistical higher-order moments and a frequency-dependent intensity correlation map. We identify correlated spectral regions within the supercontinuum associated with simultaneous sideband generation, as well as signatures of pump depletion and soliton-like pump dynamics. Experimental results are in excellent agreement with simulations.

Supercontinuum (SC) generation in optical fibers has been extensively studied over the last decade, and the underlying spectral broadening mechanisms are now well understood [1]. Attention is now increasingly focusing on the SC stability and noise properties because of important applications in source development, as well as the possibility to study links with instabilities and noise effects in other systems [2-9]. The noise properties of SC generation have of course been extensively studied using the techniques of radio-frequency RF analysis [10-11] but the recent interest in this field has been motivated by the pioneering work of Solli *et al.* who used real-time measurements of the long wavelength soliton structure of the SC to propose intriguing links with the formation of destructive rogue waves on the surface of the ocean [2]. A particular experimental originality of this work was the use of a time-stretch dispersive Fourier transform to map spectral fluctuations into temporal fluctuations that could be measured using a real time oscilloscope [12].

Real time measurements of the inherent fluctuations of fiber supercontinuum generation present significant challenges, yet there is clearly much insight to be obtained in measuring the shot-to-shot variation in spectral characteristics directly. The dispersive time stretch technique has since been applied to study aspects of supercontinuum seeding [13] and very recently to investigate the appearance of correlations between spectral modes within the sidebands generated from fiber modulation instability (MI) [14].

In this Letter, we extend previous studies of SC noise to present the first real-time characterization of noise across the complete bandwidth of a fiber supercontinuum, allowing us to directly quantify the significant differences in the observed statistical behavior with wavelength. Using dispersive time-stretching to measure spectral fluctuations in real-time, we record histograms of wavelength-dependent intensity fluctuations across the full SC spectral bandwidth, and we characterize the noise at each wavelength using statistical higher-order moments. We report results at two fiber lengths corresponding both to a broadband SC as well as the Akhmediev breather stage of evolution [4,5]. For both propagation distances, we also introduce the use of two-frequency intensity correlation maps to study fibre SC dynamics, with our results clearly showing the decrease in spectral correlation with propagation distance. This technique allows us to study correlations between widely separated spectral components of the SC, highlighting for example the simultaneous emergence of sidebands on either side of the pump, as well as pump depletion and spectral broadening dynamics. We interpret our results in terms of the fundamental nonlinear dynamics of SC propagation, and we compare our results with stochastic numerical simulations, reporting very good agreement.

**Results**

**A. Experimental Setup and Design**

We consider SC generation in the regime of noise-driven dynamics using picosecond pulses where strong shot-to-shot fluctuations are expected [1]. Our experimental setup (see Methods for full details) is shown in Figure 1(a). Picosecond input pulses at a wavelength of 1550.6 nm are amplified in a fiber amplifier (EDFA) and characterized using an optical spectrum analyser (OSA) and frequency-resolved optical gating (FROG) [15]. The pulses after amplification had temporal and spectral FWHM of 3.5 ps and 3 nm respectively. The pulses were injected into Ge-doped highly nonlinear fiber (OFS HNLF) with a coupled peak power of 70 W. The fibre zero dispersion wavelength is ~1400 nm such that there was anomalous group velocity dispersion across the full SC bandwidth generated in our experiments.

The expected fluctuations of the HNLF output preclude meaningful FROG characterization of the output field [16], but measurements using the OSA were carried out to record the average spectral characteristics observed at different propagation distances. The real-time spectral fluctuations were measured using the time stretch dispersive Fourier transform technique [12-14]. This idea behind this technique is very simple, using the fact that any pulse that propagates in a linear dispersive medium ultimately looks like its Fourier transform with sufficiently large quadratic spectral phase (see Methods). Indeed, in the spatial domain, this is one of the basic results of Fourier optics where the far-field diffraction pattern of an arbitrary spatial mask yields its spatial Fourier transform [17]. In our experiments we used a 12 GHz real time detection system and a dispersion compensating fibre (DCF) with total dispersion of +133 ps$^2$. This allowed us to measure an extensive series of single-shot spectral data to show the dramatic individual fluctuations in the noise-driven SC dynamics.

Our experimental design was informed using numerical simulations based on a stochastic generalized nonlinear Schrödinger equation (GNLSE) model in order to determine the expected propagation dynamics in the HNLF (see Methods). Although with picosecond pulse pumping we expect significant shot-to-shot variation in propagation characteristics, single-shot simulations can usefully reveal the essential dynamical features expected in this regime. Typical results are shown in Fig. 1(b) where we identify different propagation regimes: (i) an initial phase of pump self-phase modulation accompanied by spontaneous modulation instability and the appearance of well-defined spectral sidebands at a distance of ~5 m; (ii) a sideband cascade associated with high contrast pulse/breather localization [5] at a distance of ~10 m; (iii) distinct soliton emergence and spectral asymmetry with further propagation as the solitons undergo the Raman self-frequency shift. In our experiments, we characterized propagation at two different lengths of 10 m and 20 m of HNLF, the latter distance corresponding to the onset of typical SC dynamics such as soliton emergence and self-frequency shift. The spectral broadening in this case extended out to ~1700 nm with a -20 dB spectral width of around 200 nm. It was not possible to study evolution for greater propagation distances with our setup as accurate spectral measurements were not possible beyond 1750 nm.

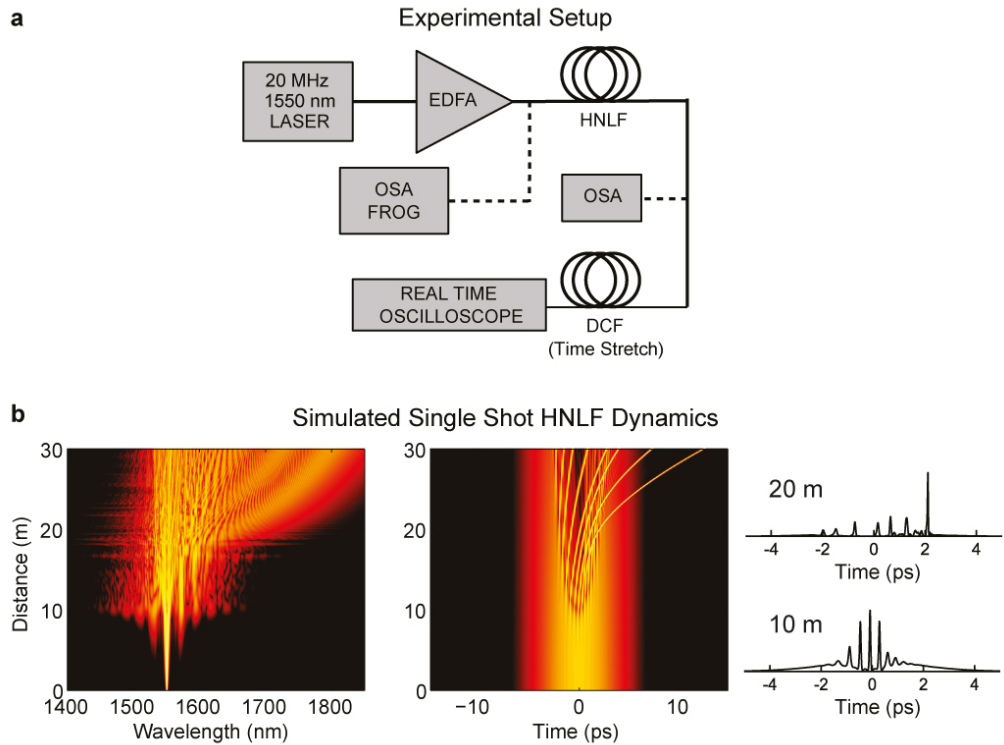

**Fig. 1** (a) Experimental setup. EDFA: Erbium doped fibre amplifier. HNLF: highly nonlinear fibre, DCF: Dispersion Compensating Fibre. OSA: optical spectrum analyser. FROG: frequency-resolved optical gating. (b) Simulated evolution of 3.5 ps pulses in HNLF as described in text showing spectral (left) and temporal (right) evolution. Temporal intensity slices at 10 m and 20 m show strong temporal localisation on the pulse envelope and soliton emergence dynamics respectively.

## B. Spectral Fluctuations and Higher-Order Moments

Our experiments recorded typically 5000 single-shot time-stretched spectra at each propagation distance. After calibration of the stretched timebase in terms of wavelength (see Methods) it is straightforward to see directly the SC fluctuations at all wavelengths in the spectra. Figure 2(a) shows the results obtained after 20 m of propagation in HNLF. The superimposed gray curves show the large shot to shot intensity fluctuations at each wavelength relative to the mean spectrum (the solid black curve). The spectral resolution here was ~1 nm. These are remarkable results, directly showing the shot to shot fluctuations in the SC, providing dramatic confirmation of the utility of the real-time capability afforded by the time-stretched dispersive transform.

We also show in the lower subfigures (b)-(d) three intensity histograms at the wavelengths shown which directly extract statistical measurements of intensity fluctuations from this data. These results clearly show the very different nature of the statistical fluctuations in different wavelength regions, with long-tailed statistics near the spectral edges and near-Gaussian statistics near the pump. Similar histogram data of this sort has of course been studied numerically [18], and previously measured in specific regimes using coarse spectral filtering [2, 12, 19-20]. Our objective here, however, is to quantify the fluctuations at high resolution across all wavelengths in the spectrum and a convenient way to do this is to analyze the intensity fluctuations in terms of the higher-order statistical moments of skewness, kurtosis and coefficient of variation [21]. In particular, with $I = I(\lambda)$ the random variable describing the intensity fluctuations at a given wavelength, the n$^{th}$-order central moment is $\mu_n = \langle (I - \langle I \rangle)^n \rangle$ with $\langle I \rangle$ the usual mean. The second order moment $\mu_2 = \sigma^2$ is the variance, and we quantify the SC noise using the coefficient of variation $C_v = \sigma^2 / \langle I \rangle$, the skew $\gamma = \mu_3 / \sigma^3$ and the kurtosis $\kappa = \mu_4 / \sigma^4 - 3$. In physical terms the coefficient of variation represents a noise to signal ratio, the skew quantifies asymmetry in the distribution, and the kurtosis is related to the presence of long-tails.

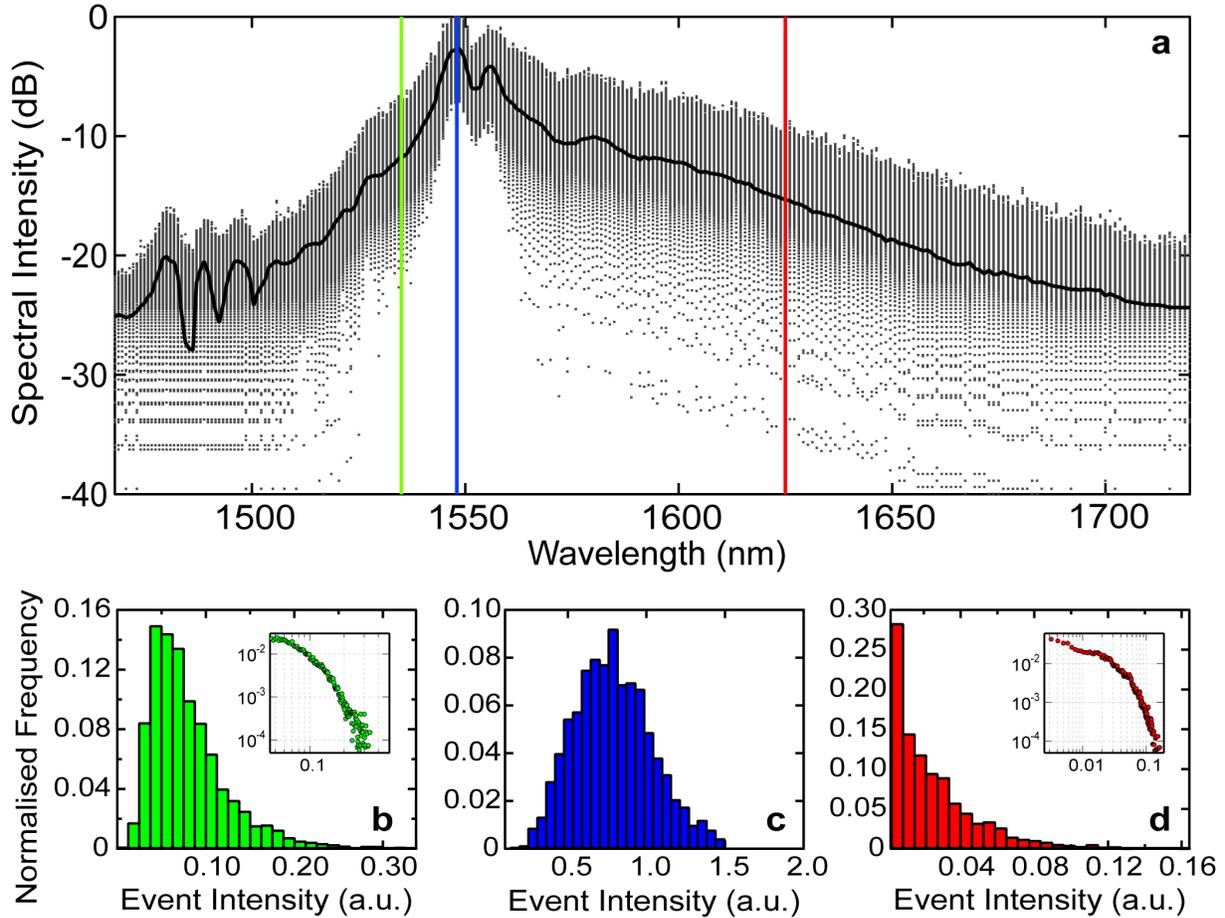

**Fig. 2** (a) Experimental results showing intensity fluctuations across the SC bandwidth after 20 m propagation in HNLF. Shot to shot realisations shown in grey; mean spectrum is the solid black line. Histogram data at wavelengths of (b) 1535 nm; (c) 1548 nm; (d) 1625 nm (these wavelengths indicated by lines of the same colour in the top figure). Histograms are centred on the wavelengths shown. Fig. 2(a) superposes 200 results obtained; the histograms are calculated from 5000 realisations.

This analysis is shown in Fig. 3, both for the SC at 20 m and also for the measured spectral data at a reduced propagation length of 10 m during the breather phase of evolution. We first present in Fig. 3(a) a visual approach to illustrate the shot-to-shot fluctuations by sequentially plotting (along the vertical direction) the results from 500 real time spectral measurements after 10 m and 20 m of propagation. The results at 10 m clearly show shot-to-shot variations, but the propagation in this regime is associated with near-symmetrical sideband structure around the pump. On the other hand, with increasing distance at 20 m, the spectral structure shows very strong asymmetry due to the effect of Raman shifting on the emergent soliton dynamics as shown in Fig. 2.

Figure 3(b) presents the quantitative analysis of the full ensemble of 5000 spectra where we plot the calculated mean spectrum (bottom subfigure) as well as the corresponding calculated statistical moments as described above (upper subfigures). Results are shown at both 10 m and 20 m as indicated. At a distance of 10 m, we see how the statistical moments in the vicinity of the pump are lower than in the sideband spectral region, illustrating how the initial pump evolution (dominated by self-phase modulation) is significantly more coherent than the process of sideband growth which is seeded by noise. As the evolution continues, however, we expect noise effects to increase also in the vicinity of the pump, and this is also seen in our experimental results at 20 m. Of particular interest here is how the results at 20 m show increasing skew and kurtosis on the long wavelength edge of the spectrum. Additional simulations allow us to identify this as being due to the spectral jitter of solitons resulting from the Raman effect. At each propagation distance, we compare experimental results with those from a series of stochastic numerical simulations (see Methods) and we see very good agreement at both propagation distances, providing confirmation of the accuracy of numerical modeling of SC generation in predicting the detailed stochastic behavior of SC generation.

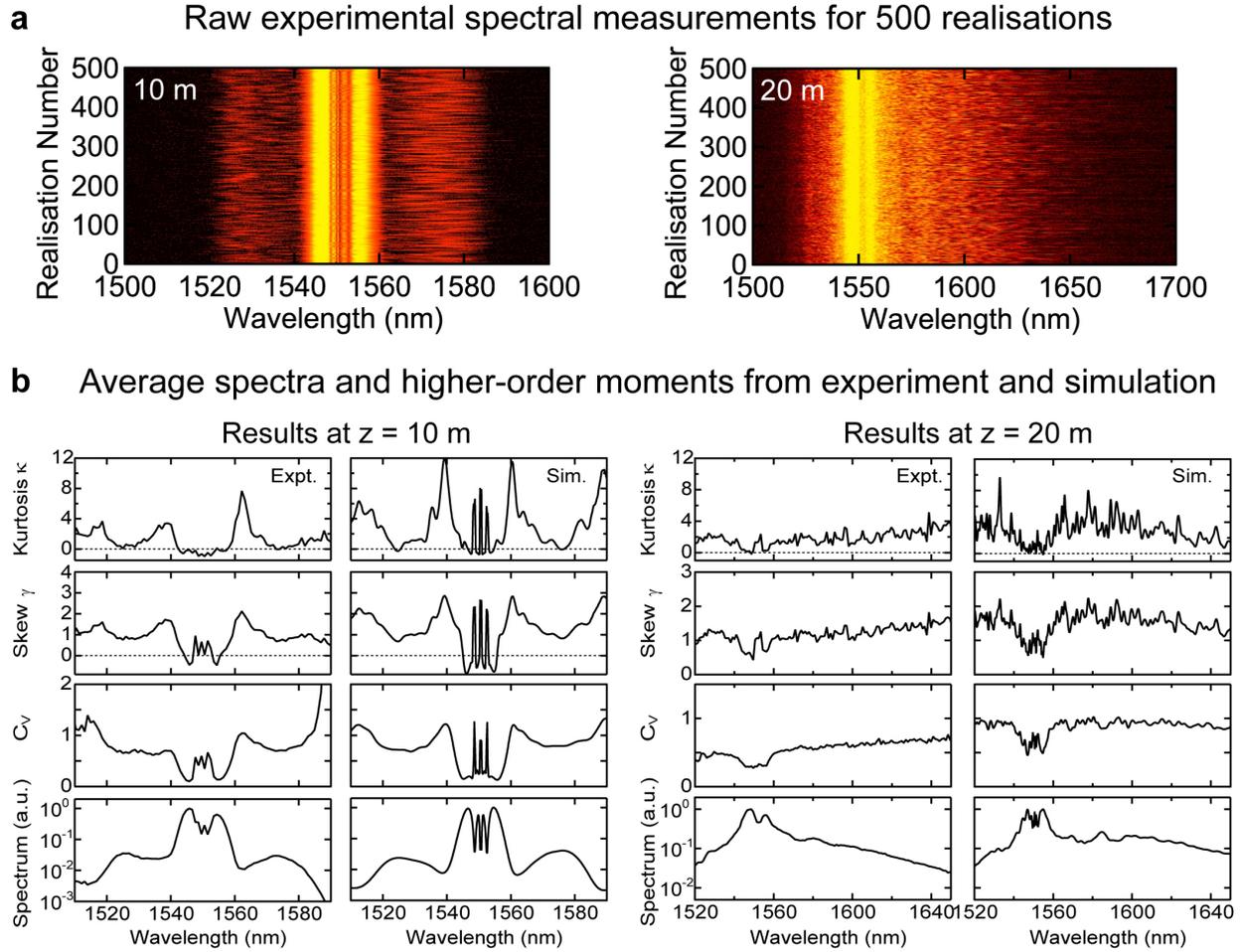

**Fig. 3** (a) False colour plot of 500 sequential real time measurements of spectra after 10 m and 20 m propagation. Note that the logarithmic colour scale used highlights the low intensity fluctuations about the pump wavelength regime which is saturated. The shot-to-shot variation in the spectral structure is very apparent. (b) Shows quantitative data over 5000 realisations plotting calculated mean spectra as well as higher order moments of Coefficient of Variation $C_v$, Skew $\gamma$ and Kurtosis $\kappa$ as indicated. At each distance both experimental (Expt) and simulation (Sim) results are shown.

## C. Intensity Correlation Maps

Higher-order moments provide a useful means of identifying regions of the SC spectrum with different noise properties. But even more insight can be obtained using an intensity correlation metric which reveals the physics of how energy is transferred between the different spectral components. Intensity correlation measurement in nonlinear fiber optics has a long history, and has been particularly motivated by studies of correlated photon pair generation for quantum information [22-24]. In a study of quantum correlations in soliton propagation [25], a frequency-domain representation was introduced to study intensity fluctuations via a two dimensional spectral correlation map, and this has also been recently used to study effects such as soliton collisions [26] and noise reduction in free-space optical filamentation from a low repetition rate source [27,28]. The recent work in Ref. 14 used real time spectral measurements to determine the inter-modal correlations of MI sidebands, but our measurements over the full bandwidth of a developing SC allow us to calculate the full bandwidth wavelength-dependent map of the intensity fluctuations. This allows to calculate not only the correlations within the components of a particular sideband, but also correlations between Stokes and anti-Stokes MI sidebands, pump-sideband correlations, and the correlation of all the other generated spectral components within the SC.

We first show numerical simulation results in Fig. 4. Here, for propagation distances in the 0-20 m range, we plot the spectral intensity correlation calculated from an ensemble of 500 simulations. The spectral correlation between two wavelengths $\lambda_1$ and $\lambda_2$ in the SC is defined as*:

$$\rho(\lambda_1, \lambda_2) = \frac{\langle I(\lambda_1)I(\lambda_2)\rangle - \langle I(\lambda_1)\rangle\langle I(\lambda_2)\rangle}{\sqrt{\left(\langle I^2(\lambda_1)\rangle - \langle I(\lambda_1)\rangle^2\right)\left(\langle I^2(\lambda_2)\rangle - \langle I(\lambda_2)\rangle^2\right)}}, \quad (1)$$

where $I(\lambda)$ is the array of intensities at a particular wavelength $\lambda$ in the supercontinuum from the ensemble, and the angle brackets represent the average over the ensemble. The spectral correlation varies over the range $-1 < \rho < 1$ with the colour scale in the figure indicating the degree of correlation. Positive correlation $\rho\,(\lambda_1, \lambda_2) > 0$ (shades of red) indicates that the intensities at the two wavelengths $\lambda_1, \lambda_2$ increase or decrease together, and of course we trivially observe perfect correlation $\rho\,(\lambda_1, \lambda_2) = 1$ (yellow) along a positive diagonal line when $\lambda_1 = \lambda_2$. Negative correlation or anti-correlation $\rho\,(\lambda_1, \lambda_2) < 0$ (shades of blue) indicates that as the intensity at one wavelength e.g. $\lambda_1$ increases, that at $\lambda_2$ decreases and vice-versa. The correlation function is also symmetric across the positive diagonal such that $\rho\,(\lambda_1, \lambda_2) = \rho\,(\lambda_2, \lambda_1)$.

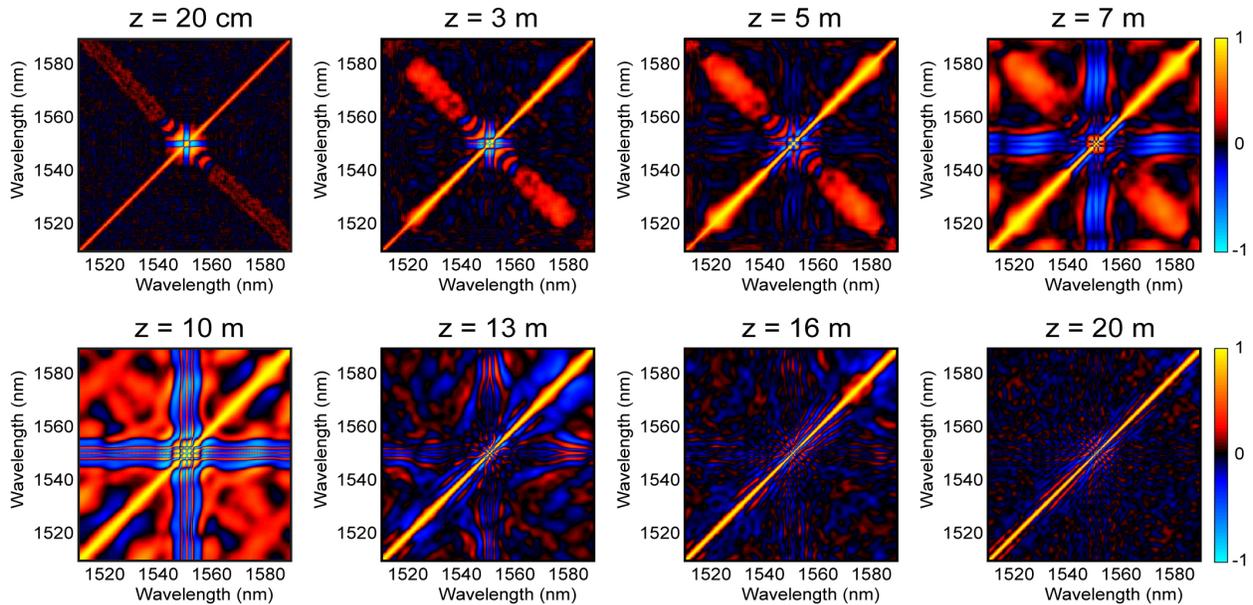

**Fig. 4** Simulation results showing the evolution of the intensity correlation map $\rho\,(\lambda_1, \lambda_2)$ with propagation distance. The colour scale is shown on the right and is such that positive correlations are yellow, negative correlations are blue and black represents the absence of correlation. The traces are centred on the pump wavelength at 1550.6 nm.

The results in Fig. 4 allow us to see clearly how particular physical processes that influence the field evolution at different distances introduce different signatures into the correlation map. This is apparent even after 20 cm of propagation when we see a developing red negative diagonal associated with strong positive correlation ($\rho > 0$) between conjugate MI wavelengths about the pump. This arises from the simultaneous growth of sideband pairs as the pump induces symmetric sideband generation through modulation instability from noise [29]. The map at 20 cm also shows a different class of evolution associated with a negative correlation in the form of a blue cross-like structure ($\rho < 0$) in the centre around the pump. This indicates that radiation at the pump wavelength is anti-correlated with radiation in adjacent spectral regions in the range 1540-1550 nm and 1551-1560 nm. This arises from intra-pulse energy exchange dynamics as the input pulse evolves similar to a higher-order soliton, undergoing spectral broadening and transferring energy to new spectral components in the wings [25-26]. Signatures of sideband generation (positive correlation) and soliton-like dynamics (negative correlation) continue to develop with propagation up to ~10 m. The fine modulation on the correlation structure of the pump is due to self-phase modulation [25].

We also see how the correlated spectral regions along both diagonals broaden with propagation as the overall spectral bandwidth increases. Propagation above 10 m also sees increasing noise transfer between the different spectral regions of the developing SC, and it is clear how after ~13 m, noise effects begin to dominate any correlation signatures as the spectrum becomes increasingly incoherent. We see essentially zero correlation everywhere aside from the trivial $\lambda_1 = \lambda_2$ diagonal and a small area around the pump corresponding to the residual coherent self-phase modulation.

The decrease in correlation with distance is a well-known effect in noise-driven SC generation which can arise both from spontaneous MI when using long pulse excitation, and from noise-induced soliton fission dynamics with femtosecond pulses [1]. In the specific regime of SC generation considered here, we see decoherence both from MI and soliton dynamics. The effect of pump noise first induces fluctuations in spectral amplitude and phase close to the pump frequency due to the initial growth of MI sidebands. With increasing propagation, the noise and decoherence then increases due to jitter on the sub-picosecond solitons emerging from the strong (but noisy) modulation on the pulse envelope which was induced by the initial MI.

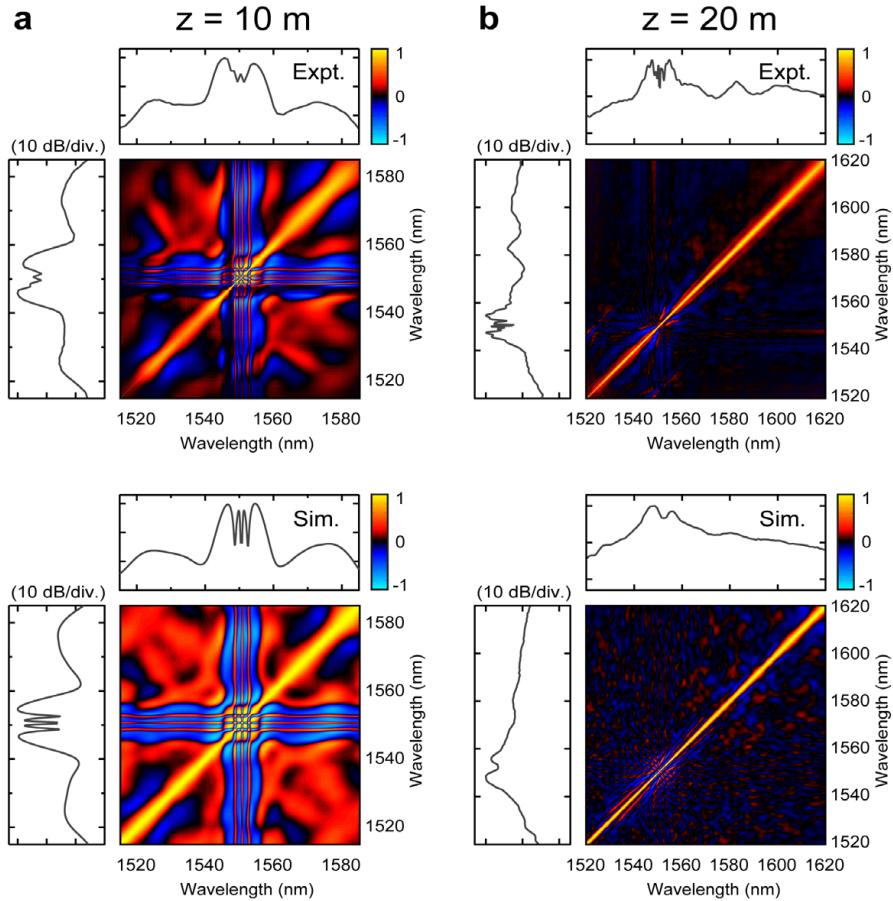

**Fig. 5** For propagation in (a) 10 m and (b) 20 m of HNLF, the results show experimental (top) and simulation (bottom) correlation maps $\rho(\lambda_1, \lambda_2)$ as described in the text.

The simulation results at 10 m and 20 m have been confirmed experimentally as shown in Fig. 5 (a) and (b). Here we also reproduce the simulation results in the lower subfigures and project spectra onto each axis to aid in interpretation of the spectral features. Of particular note in interpreting these results is how the experimental data accurately shows the complex wavelength-dependent spectral correlation expected from simulation. The experiments also clearly show the decoherence from 10 m to 20 m propagation as the degree of spectral correlation approaches zero across the SC.

**Discussion**

Supercontinuum generation in optical fibre is a rich and complex process that involves multiple nonlinear effects that influence spectral broadening in different ways. For the particular case of excitation by picosecond pulses as considered here, the evolution consists of an initial phase of noise-seeded modulation instability that precedes the development of more strongly-localized temporal breather structures and emergent solitons. The development of real-time measurement techniques of supercontinuum noise has already provided insights into optical rogue wave soliton emergence and inter-sideband correlations in modulation instability and our results here have shown how it can be applied to study the full wavelength-dependence of noise across the full SC spectrum. To our knowledge these results represent the first experimental characterisation of the shot-to-shot SC fluctuations across the full spectral range and the first comparison of the intensity correlation map to characterise and interpret SC propagation dynamics. In fact, it is clear that the metrics of higher-order moments and the intensity correlation map reveal detailed information about the particular dynamics of nonlinear propagation and noise effects that are not apparent in averaged measurements and are highly complementary to other techniques for noise characterisation such as RF spectral analysis.

We also stress the simplicity of the dispersive time-stretch technique and the straightforward way in which the statistical and correlation metrics can be extracted from experimental data. We suggest that this technique of real-time spectral noise characterisation should now become the standard approach to analysing noise effects in nonlinear fibre propagation as it provides a powerful means of allowing rapid experimental optimization, and generating large data sets from experiment in this way is orders of magnitude more efficient than stochastic simulations. Indeed, we anticipate wide application of this technique for detailed studies of noise in broadband pulse propagation, the characterization of quantum-optical intensity-correlations in soliton dynamics, and in the study of spectral instabilities in ultrafast nonlinear optics in general.

**Methods**

Our experiments used a Pritel FFL picosecond laser and Keopsys BT-Series EDFA to generate the amplified input pulses to the HNLF. The source repetition rate was 20 MHz. Averaged spectral measurements used an Anritsu MS9710B OSA with 0.07 nm spectral resolution. We implemented dispersive-time stretching for spectral measurements using a high speed real-time oscilloscope (Tektronix 40 GSamp/sec, 12 GHz bandwidth) and a New Focus 20 GHz photodiode.

Dispersive time stretching is described briefly as follows. For a temporal field $U(t)$ with Fourier transform $\tilde{U}(\omega)$, propagation in a length $z$ of "stretching" fiber of dispersion $\beta_{2S}$ yields (for large $\beta_{2S} z$ in a stationary phase approximation) a temporally-dispersed output pulse of the form: $U_z(t) \sim \exp(-it^2/2\beta_{2S}z)\tilde{U}(t/\beta_{2S}z)$ such that $|U_z(t)|^2 \sim |\tilde{U}(t/\beta_{2S}z)|^2$ [30]. In other words, the intensity of the dispersed output pulse yields the pulse spectrum subject to the simple mapping of the time coordinate of the stretched pulse to frequency $v$ (in Hz) where $2\pi v = t/\beta_{2S}z$. In our experiments, we used 875 m of DCF (Sumitomo Electric Industries) with total normal dispersion of $\beta_{2S}z = +133$ ps$^2$. We attenuated the input to the DCF in order to ensure linear propagation, and confirmed the fidelity of the time-stretching technique using numerical simulations including higher-order dispersion, nonlinearity and Raman in the DCF and by comparing the OSA spectra and the averaged time-stretched spectrum. The typical duration of the time-stretched pulse is ~ 10 ns at the -20 dB level such that with the 50 ns pulse period of our source laser, we could record time series of up to 8000 spectra using an extended sweep and the available oscilloscope storage capacity. The equivalent 0.8 nm spectral resolution of our setup was determined by the 12 GHz oscilloscope bandwidth.

An alternative approach to data acquisition used delayed triggering to isolate and acquire the stretched time trace over a restricted temporal window, acquiring only one stretched pulse at a time, or a selected portion of the pulse, over a limited equivalent wavelength range. The latter approach is a particularly visual way to see the dramatic shot-to-shot fluctuations. Moreover, it reduces the effect of instrumental noise on measured statistics by dynamic gain control at different trigger delays to ensure that the fluctuations to be measured were always much larger than any noise floor over any wavelength range of interest.

Our numerical simulations used a well-known generalized NLSE model described in Ref. 1. The use of a broadband EDFA precluded the use of a rigorous phase-diffusion model [31] for the simulation input noise field in our simulations but we found good agreement between experiment and simulation with a random phase spectral background field at the –40 dB level relative to the peak of the pump pulse. This noise level corresponded to that introduced by the EDFA measured experimentally. The nonlinearity and dispersion parameters of the HNLF at the pump wavelength are: $\gamma$ = 9.4 x $10^{-3}$ $W^{-1}$ $m^{-1}$, $\beta_2$ = -5.239 $ps^2$ $km^{-1}$, $\beta_3$ = 4.290 x $10^{-2}$ $ps^3 km^{-1}$. Simulations were performed including terms to $\beta_3$ in the dispersion and higher-order nonlinear effects of self-steepening and Raman scattering. The coupled peak power of 70 W into the HNLF was calculated based on measurements of the average output power from the HNLF.

## Acknowledgements

We acknowledge support from the French Agence Nationale de la Recherche (ANR-09-BLAN-0065 IMFINI), the Academy of Finland Research grants 132279 and 130099, and the European Research Council Advanced Grant MULTIWAVE.

## Author Contributions

B.W., A.S, L.L., P.A.L., J.M.M. and J.M.D. designed and performed the experiments. The experimental design, detection setup fidelity was checked and independently validated by T. S., A. M. and A. K. Simulations were developed and implemented by B.W., J.M.D., G.G. and F.D. All authors participated in the analysis and interpretation of the results and the writing of the paper.

## Competing Financial Interests Statement

The authors declare that they have no competing financial interests.